\documentclass[12pt]{article}
\usepackage{amsmath, amssymb}
\usepackage{graphics}
\usepackage{epsfig}
\newcommand{\PRA}[1]{Phys.\ Rev.\ A {\bf #1}}

\newcommand{\JPA}[1]{J.\ Phys.\ A {\bf #1}}

\newcommand{\JTB}[1]{J.\ Theor.\ Biol. {\bf #1}}
\newcommand{\PRSB}[1]{Proc.\ Roy.\ Soc.\ B. {\bf #1}}
\newcommand{\PTP}[1]{Prog.\ Theo.\ Phys. {\bf #1}}
\newcommand{\IJMPC}[1]{Int.\ J.\ Mod.\ Phys.\ C {\bf #1}}
\begin{document}

\title{\textbf{Lattice three-species models of the spatial spread of rabies
among foxes}\\}
\author{\textbf{A.~Benyoussef$^1$,~N.~Boccara$^{2,3}$,~H.~Chakib$^1$}\\
   \textbf{and~H.~Ez-Zahraouy$^1$}\\
\\
$^1$ Laboratoire de Magn\'etisme et Physique des Hautes\\
\'Energies, D\'epartement de Physique, Facult\'e des Sciences,\\
B.P. 1014, Rabat, Morocco.\\
$^2$ DRECAM/SPEC, Centre d'\'Etude de Saclay,\\
91191 Gif-sur-Yvette Cedex, France.\\
$^3$ Department of Physics, University of Illinois, Chicago,\\
IL 60607-7059}

\maketitle

\begin{abstract}

Lattice models describing the spatial spread of rabies among foxes are
studied. In these models, the fox population is divided into
three-species: susceptible (S), infected or incubating (I), and infectious
or rabid (R). They are based on the fact that susceptible and incubating
foxes are territorial while rabid foxes have lost their sense of direction
and move erratically. Two different models are
investigated: a one-dimensional coupled-map lattice model, and a
two-dimensional automata network model. Both models take into account
the short-range character of the infection process and the diffusive motion
of rabid foxes. Numerical simulations show how the spatial distribution of
rabies, and the speed of propagation of the epizootic front
depend upon the carrying capacity of the environment and diffusion of
rabid foxes out of their territory.

\end{abstract}

\vspace{2cm}

\maketitle

\newpage

\section{Introduction}

Rabies is one of the oldest recorded infectious diseases. Records
from the Middle Ages show that rabies was then widespread in western
Europe. Throughout the world, dogs are important vectors, however,
there exist many early references to several species of wildlife
implicated in the spread of the disease~\cite{baer}, \cite{sike},
\cite{mcdo}. Nowadays, most
cases of rabies are found among animals (dogs, cats, livestock,
wildlife). According to Macdonald~\cite{mcdo}, the present European
epizootic (epidemic among animals) began south of Gdansk (Poland)
in 1939. Moving westwards, it was first recorded in France in 1968,
the Netherlands in 1974, Spain in 1975 and Italy in 1977. Its main
vector and victim is the red fox.

Rabies is an acute infectious disease of the central nervous
system caused by a virus. The disease is transmitted from rabid to
susceptible foxes, usually by biting. About half of the infectious
foxes become agressive, and lose their sense of direction and
territorial behavior, wandering randomly.

Murray and co-workers~\cite{kall}, \cite{msb}, \cite{mur} studied
different models of the spread
of rabies among foxes formulated in terms of partial differential
equations. In these models, the fox population is divided into
three species:
susceptible, infected but non-infectious, and rabid
foxes which are infectious. The principal assumptions are:
\begin{enumerate}
\item[(i)] Susceptibles evolve in time according to a logistic
equation.
\item[(i)] Susceptibles become infected at an average rate per
capita proportional to the number of rabid foxes present.
\item[(iii)] Infected foxes become rabid after an average
incubation period.
\item[(iv)] Rabid foxes die after an average duration of the
disease.
\item[(v)] Rabid and infected foxes also die of causes other than
rabies.
\item[(vi)] The spatial spread of the disease is essentially due
to the random motion of the rabid foxes
\end{enumerate}

Such models have unquestionably contributed to our understanding of
the spread of an infectious disease, but they do not take
correctly into account the short-range character of the infection
process, and neglect spatial correlations. In order to include
these features, we have built up and studied two lattice models in which the
spread of the epidemic is viewed as the growth of a random cluster
on a lattice. We will first describe a coupled-map lattice model, and
then a two-dimensional automata network model.

\section{Lattice models}

The spread of an epidemic in a population is a complex process. When
modeling such a process, among the many features which are likely to
be important, we should, however, only retain the few relevant ones
which are thought to play an essential role in the interpretation of
the observed phenomena. In both models, at each time step, the fox
population evolves according to the following rules:
\begin{enumerate}
\item[(1)] A susceptible has a probability $b$ to give birth
to a susceptible at a nearby empty location. Infected and rabid
foxes do not give birth.
\item[(2)] Susceptibles and infected have a probability $d$ to
die due to natural causes, and $d_\ell$ multiplied by the total local
population density of neighboring foxes (\textit{i.e.}\ susceptibles
+ infected + infectious) due to lack of food.
\item[(3)] Susceptibles become infected at a rate proportional to
the local density of neighboring rabid foxes, the proportionality
factor is the probability $p_i$ to be infected.
\item[(4)] Infected, \textit{i.e.} incubating, foxes become rabid
with a probability $p_r$.
\item[(5)] Non-rabid foxes being territorial, susceptibles and
infected foxes evolve without moving to a neighboring territory.
\item[(6)] Rabid foxes have a probability $d_r$ to die due to the
disease, and $d_\ell$ multiplied by the total local density of
neighboring foxes due to lack of food.
\item[(7)] Rabid foxes move at random to a neighboring location.
\end{enumerate}

In order to exhibit some realistic features of the spread of a rabies
epidemic, following the analysis of Murray \textit{et al\/}~\cite{msb},
our models have to contain a minimum number of parameters.
There is one source term coming from the birth of
susceptibles (parameter $b$). To account for the three death processes,
natural causes, lack of food, and rabies, we need 3 parameters ($d$,
$d_\ell$, and $d_r$). We need to introduce a parameter which measures
the probability to be infected by contact (parameter $p_i$). The
existence of an incubation period is an essential feature to exhibit
decreasing periodic fluctuations following the main wave front of
the susceptible population (parameter $p_r$).

\subsection{Coupled-map lattice model}

A coupled-map lattice is a dynamical system in which space and time are
discrete variables while states are continuous~\cite{kane},
\cite{wk}.
Here the state of the system is represented by the function
$\mathbf{P}:(i,t)\mapsto \mathbf{P}(i,t)$, where
$(i,t)\in\mathbb{Z}\times\mathbb{N}$, and $\mathbf{P}\in[0,1]^3$ is a
three-dimensional vector whose components are $S(i,t)$, $I(i,t)$,
and $R(i,t)$ which denote, respectively, the densities at site
$i$ and time $t$ of susceptible, infected and rabid foxes.
$\mathbb{Z}$ is the set of all integers and $\mathbb{N}$ is the
set of nonnegative integers. In this model, each site corresponds
to a specific territory. Susceptibles and infected foxes evolve
without moving to a neighboring site. On the contrary, rabid
foxes, which have lost their sense of direction, move to one of
their two neighboring sites with equal probabilities . According
to our assumptions (Rules 1 to 7), the dynamics of the system is governed by the
following recurrence relations:
\begin{align*}
S(i,t+1) &=  S(i,t) - d S(i,t) - d_\ell N(i,t)S(i,t)\\
         &\qquad + b (1-N(i,t))S(i,t)- p_i R(i,t)S(i,t)\\
I(i,t+1) &=  I(i,t) - d I(i,t) - d_\ell N(i,t)I(i,t)\\
         &\qquad + p_i R(i,t)S(i,t) - p_r I(i,t)\\
R(i,t+1) &=  R(i,t) - d_r R(i,t) - d_\ell N(i,t)R(i,t) + p_r I(i,t)\\
         &\qquad + D(R(i-1,t)) + R(i+1,t) - 2 R(i,t)),
\end{align*}
where $N(i,t)=S(i,t)+I(i,t)+R(i,t)$ is the total fox density at
site $i$ and time $t$, and $D$ is the diffusion coefficient
characterizing the random motion of the rabid foxes.

\subsection{Automata network model}

An automata network is a fully discrete dynamical system. That is,
space, time and states are discrete variables. In our model, the space
consists of a square $L\times L$. Since we are interested in the spread
of an epidemic starting a the center of the lattice, we did not
choose cyclic boundary conditions. As for a coupled-map lattice,
the time variable is a nonnegative integer. The state of the system at
time $t$ is described by the function $s_t:(i,j)\mapsto s(i,j;t)$,
where $(i,j)\in{\mathbb{Z}}_L^2$, $t\in\mathbb{N}$, and $s(i,j;t)$
can take four values corresponding to the four possible states of
the site $(i,j)$ since this site can be either empty or occupied
by one of the three fox species (susceptible, infected, rabid).
At each time step, the state of the system evolves according to
the successive application of the two following subrules:
\begin{itemize}
\item first a local rule describing birth, death and infection
processes, which is applied to all sites synchronously;
\item then, a motion rule mimicking rabid fox erratic motion.
\end{itemize}

Subrule (1) is a probabilistic, two-dimensional, four-state
cellular automaton rule, which is defined as follows. At each time step,
\begin{enumerate}
\item[(i)] each susceptible has a probability $b$ to give birth to
a susceptible at an empty first-neighbor site, therefore, the
probability that an empty site, having $z_s$ susceptibles among its
four first-neighbors, becomes occupied by a susceptible is
$1-(1-b)^{z_s}$;
\item[(ii)] each susceptible has a probability $d$ to die of natural causes;
\item[(iii)] each susceptible having $z_i$ rabid foxes among its
four first-neighbors becomes infected with a probability
$1-(1- p_i)^{z_i}$;
\item[(iv)] each infected has a probability $d$ of die of natural causes;
\item[(v)] each infected has a probability $p_r$ to become
rabid;
\item[(vi)] a rabid fox has a probability $d_r$ to die  of the disease;
\item[(vii)] each fox (either susceptible, or infected, or rabid) having
$z_f$ foxes among its first-neighbors has a probability $d_\ell z_f$ to die
due to lack of food.
\end{enumerate}

Subrule (2) can be described as follows:
At time $t$, a rabid fox is selected at random to perform a move to a
first-neighbor site also selected at random. If the site is empty the
fox will effectively move otherwise it will not. This process is repeated
$m N_R(t) L^2$, where $N_R(t)$ denotes the density of rabid foxes at time $t$.
The parameter $m$ represents the average number of tentative move
per rabid fox at time $t$. Note that this sequential diffusive process allows
some rabid foxes to move more than others.

Automata networks of this type are called \textit{diffusive cellular
automata}. Their general properties have been studied by Boccara
\textit{et al\/}~\cite{boc1}. They have been used to build up various
epidemic models~\cite{boc2} in which the motion of the individuals play
an important r\^ole in the spread of the epidemic.

\section{Numerical simulations: results and analysis}

\subsection{Coupled-map lattice model}

In our simulations, the lattice size is 441. The sites are labeled
from $-220$ to $220$. The initial susceptible fox densities have the same value
at all sites $i$, while the initial infected and rabid fox densities are
nonzero only at the central site. These values are: $S(i,0)=0.6$.
$I(0,0)=0.005$, and $R(0,0)=0.005$. For all $t\in\mathbb{N}$, at the
boundaries of the chain, the densities of the various fox species satisfy
the condition
\begin{align*}
S(L+1,t) + &I(L+1,t) + R(L+1,t) = \\
           &S(-L-1,t) + I(-L-1,t) + R(-L-1,t) = 0
\end{align*}

As reported by Macdonald~\cite{mcdo} and Murray~\cite{mur}, the epidemic
spread of rabies among foxes is characterized by a traveling epizootic wave
front followed by periodic decreasing fluctuations of susceptibles
density which tends to its steady state. Our numerical simulations
show that our coupled-map lattice model clearly exhibits these features,
as illustrated in Figure~1 which represents, at a
given time, the variations of susceptible, and rabid fox densities,
as a fonction of site location for two different values of $d_\ell$.
Increasing $d_\ell$ decreases the height of the peak of the first and
subsequent outbreaks of infected and rabid foxes. A smaller
value of $d_\ell$ is equivalent to a larger carrying capacity of
the environment, therefore, a larger carrying capacity implies a
more severe epidemic.

Figure~2 illustrates the influence of the diffusion
coefficient $D$. As expected, increasing $D$ increases the speed at
which the epidemic spreads (see also Figure~3).
While the values of the amplitudes of the
various fox densities do not change, the distance between two
successive outbreaks increases with $D$.

According to epidemiological evidence (cf. references in Murray~\cite{mur})
there exists a threshold value for the carrying capacity below which
rabies die out. In the case of our model we found that for
$d_\ell>0.0456$ the epizootic does not spread (see Figure~4).
This threshold depends, of course, upon the values of the other parameters.

We have determined numerically the speed of the epizootic wavefront as a function
of the diffusion coefficient $D$ for two different values of the parameter
$d_\ell$. Our results are represented in Figure~3. As expected from a dimensional
argument, this speed varies as $\sqrt{D}$. Increasing $d_\ell$,
we verify again that the speed of the epizootic wavefront decreases.

We have also determined numerically the speed of the epizootic wavefront
as a function of the parameter $d_\ell$. Figure~4
shows that, as we already mentioned, this speed is a decreasing function
of $d_\ell$, which goes to zero at a threshold value.

\subsection{Automata network model}

In our simulations, the lattice size is $201\times 201$.
Simulations start from a random initial configuration. The initial
densities of susceptible, infected and rabid foxes are, respectively,
equal to $0.6$, 0.005, and 0.005.
Rabid and infected foxes exist only inside a disk of radius
10 in the initial configuration. Our results are averages over
50 to 100 different initial configurations.

As for the coupled-map lattice model, we have studied the influence
of the carrying capacity and the diffusion on the various fox species.
We found similar results as illustrated in Figures~5a-5c. Increasing
$d_\ell$ decreases the height of the peak of the first and subsequent
outbreaks of rabid foxes (Figures~5a and 5b), while increasing $m$
increases the speed at which the epidemic spreads (Figures~5b and 5c).
Above a threshold value of $d_\ell$, which is equal to 0.053 for $m=0.5$
and 0.0245 for $m=0.1$, the epizootic does not spread (see Figure~9).

As time increases, rabid foxes can be found at larger distances
from the center of the lattice. For different values of the parameter
$m$, we have determined the fractal dimension of the
rabid cluster---that is the cluster containing all rabid foxes---as a
function of time. If $r_{\max}$ is the radius of this cluster, its
fractal dimension $\delta$ is defined by $N_R(t)=\pi r_{\max}^\delta$,
where $N_R(t)$ is the number of rabid foxes in the cluster.
Figure~6 shows that $\delta$ tends to a constant value as time increases.
This limit value increases when $m$ decreases since, as expected,
the density of rabid foxes if higher for lower values of $m$. For
$d_\ell$ respectively equal to 0.005, 0.01, and 0.015, the
corresponding limit values are 1.27, 1.25, and 1.22.

Figures~7a-7f show how the cluster containing the infected and infectious
foxes grows as a function of time. Infected and infectious foxes are essentially
localized at the boundary of the cluster. This feature corresponds to the
small subsequent outbreaks following the first one. This is also clearly
illustrated in Figures~5.

As for the coupled-map lattice model, we have determined numerically how
the speed of the epizootic wavefront varies with the carrying capacity and
the diffusion. Our results are represented in Figures~8 and 9. The curves
representing the variation of the speed of the epizootic wavefront
as a function of $d_\ell$ show the existence of a threshold value,
which, in particular, depends upon the parameter $m$.

\section{Conclusion}

We have investigated two different models of the spatial spread of
rabies among foxes: a one-dimensional coupled-map lattice model, and a
two-dimensional automata network model. In both models, the fox population
is divided into three-species: susceptible, infected or incubating, and
infectious or rabid. They are based on the fact that susceptible and incubating
foxes are territorial while rabid foxes have lost their sense of direction
and move erratically out of their territory propagating the disease.
coupled-map lattice models and automata networks models have the
advantage, compared to models formulated in terms of differential
equations, to take into account the short-range character of the
infection process. We have essentially studied how the spatial distribution of
rabies, and the speed of propagation of the epizootic front
depend upon the carrying capacity or food availability of the environment
and parameters characterizing the erratic motion of rabid foxes out of
their territory. In agreement with ecological studies, our numerical
simulations show that, decreasing food availability slows down the spread
of the disease, and, that below a certain threshold, rabies eventually dies
out. On the other hand increasing the parameters measuring the
diffusive motion of rabid foxes favors the spread of the disease.

\newpage

\section*{Figure captions}

{\parindent = 0pt

Figure 1- \textit{Susceptible (S) and rabid (R) densities as
functions of site location at time $t=1800$ for $b=0.01$,
$d=0.001$, $d_r=0.1$, $p_1=0.8$, $p_r=0.05$, $D=0.1$,
$d_\ell=0.01$ (continuous line), and $d_\ell=0.015$ (dashed
line).}

Figure 2- \textit{Fox populations densities as a function of site location
at time $t=1800$ for $b=0.01$, $d=0.001$, $d_\ell =0.01$, $d_r=0.1$,
$p_1=0.8$, $p_r=0.05$, $D=0.1$ (dashed line), and $D=0.2$ (continuous line).}

Figure 3- \textit{Speed of the epizootic wavefront as a function of
$D$ for $b=0.01$, $d=0.001$, $d_r=0.1$, $p_1=0.8$, $p_r=0.05$, and two
different values of $d_\ell$ indicated in the figure.
These two set of data can be fitted with functions of the form
$A\sqrt{D}$ (continuous lines)}

Figure 4- \textit{Speed of the epizootic wavefront as a function of
$d_\ell$ for $b=0.01$, $d=0.001$, $d_r=0.1$, $p_1=0.8$, $p_r=0.05$, and $D=0.1$.}

Figure 5- \textit{Fox populations densities as a function of site location
at time $t=1600$ for $b=0.01$, $d=0.001$, $d_r=0.1$, $p_i=0.8$, $p_r=0.05$,
and (a) $d_\ell=0.01$, $m=0.1$, (b) $d_\ell=0.015$, $m=0.1$, (c) $d_\ell=0.015$,
$m=0.2$.}

Figure 6- \textit{Fractal dimension of the rabid cluster as a function of
time for $b=0.01$, $d=0.001$, $d_r=0.1$, $p_i=0.8$, $p_r=0.05$, $m=0.2$ and
different values of $d_\ell$ indicated in the figure.}

Figures 7- \textit{Spatial patterns obtained after different numbers of time
steps. The symbols: {\Huge{$\cdot$}}, $\triangle$, and $\ast$ denote,
respectively, susceptible, infected and rabid foxes. Parameters
values: $b=0.01$, $d=0.001$, $d_l=0.001$, $d_r=0.1$, $p_i=0.8$, $p_r=0.05$,
$m=0.2$. (a) $t=0$, (b) $t=50$ (c) $t=100$, (d) $t=200$, (e) $t=500$, (f) $t=800$.}

Figure 8- \textit{Speed of the epizootic wavefront as a function of $m$ for
$b=0.01$, $d=0.001$, $d_r=0.1$, $p_i=0.8$, $p_r=0.05$, and two different values of
$d_\ell$ indicated in the figure.}

Figure 9- \textit{Speed of the epizootic wavefront as a function of $d_\ell$ for
$b=0.01$, $d=0.001$, $d_r=0.1$, $p_i=0.8$, $p_r=0.05$, and two different values of
$m$ indicated in the figure.}

}

\begin{figure}[htbp]
\epsfig{figure=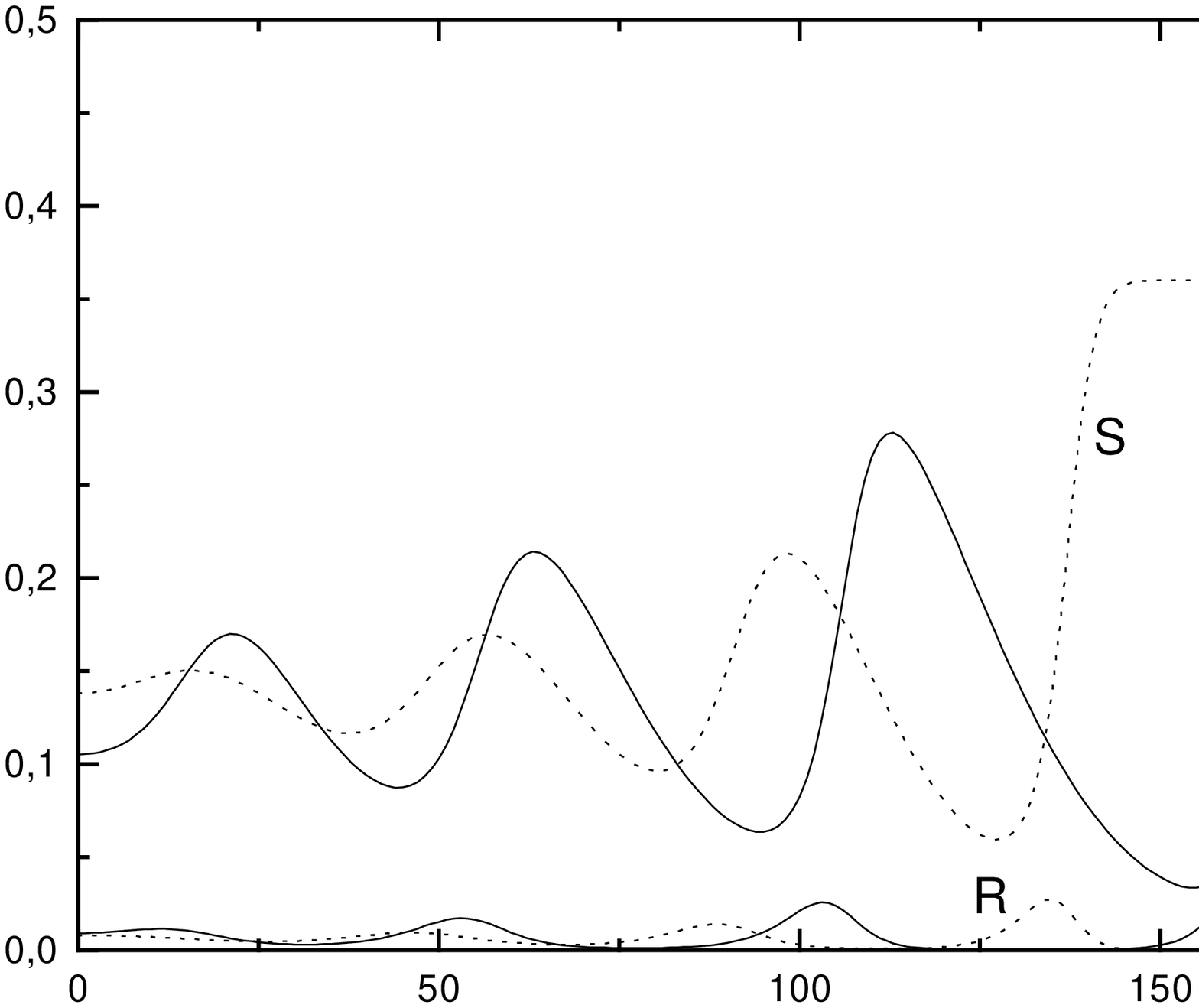, scale=0.6}
\end{figure}
\begin{figure}[htbp]
\epsfig{figure=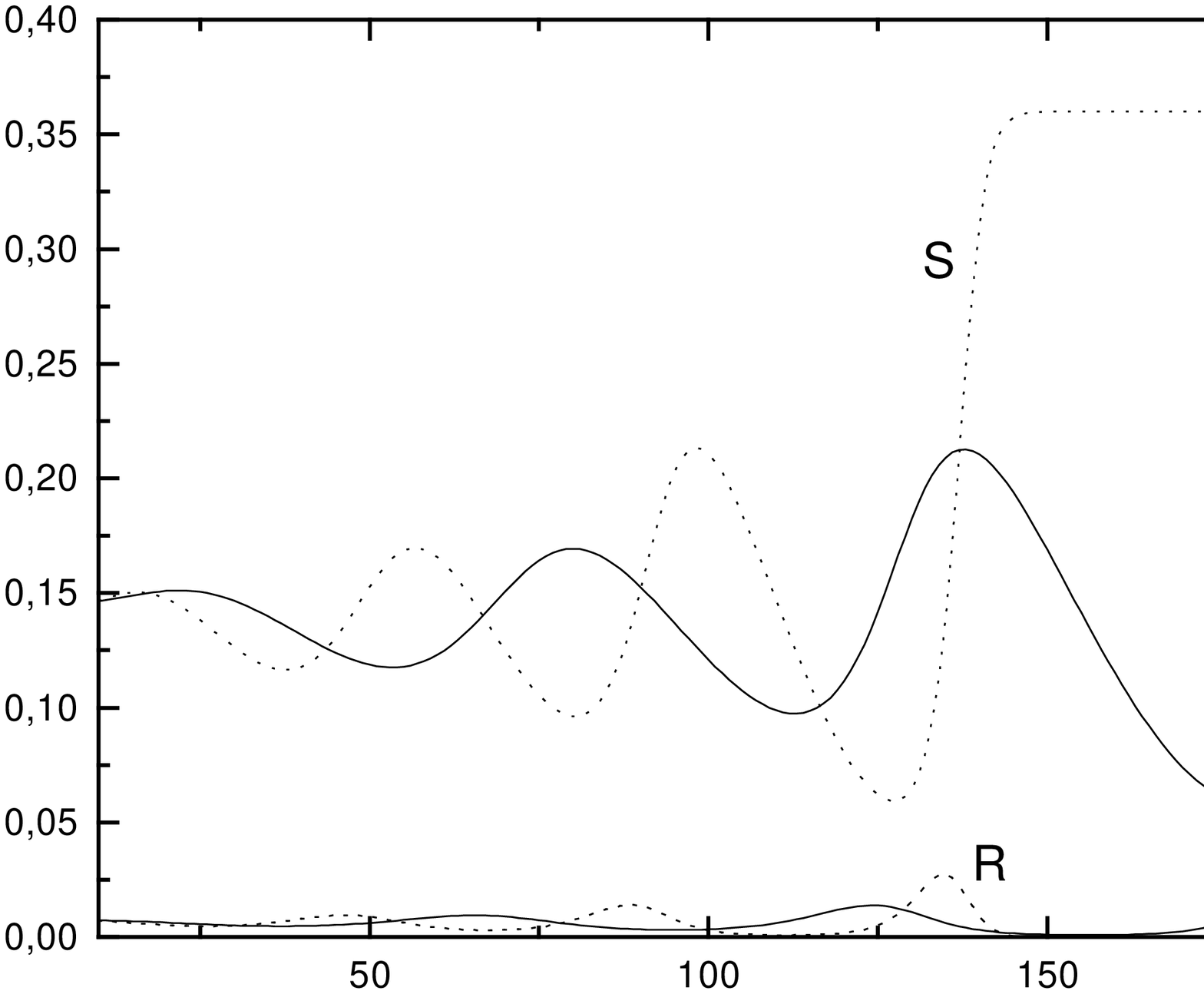, scale=0.6}
\end{figure}
\begin{figure}[htbp]
\epsfig{figure=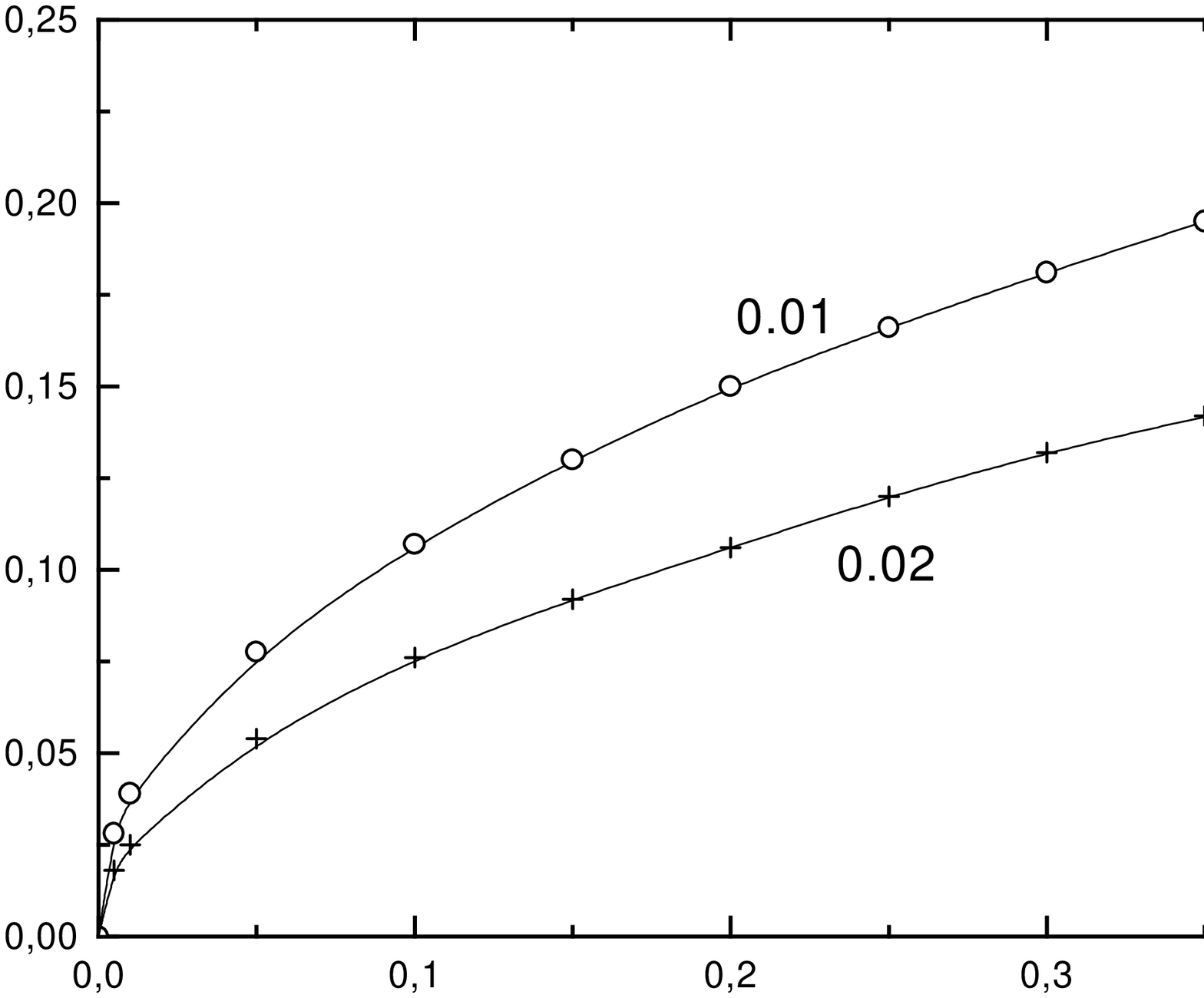, scale=0.6}
\end{figure}
\begin{figure}[htbp]
\epsfig{figure=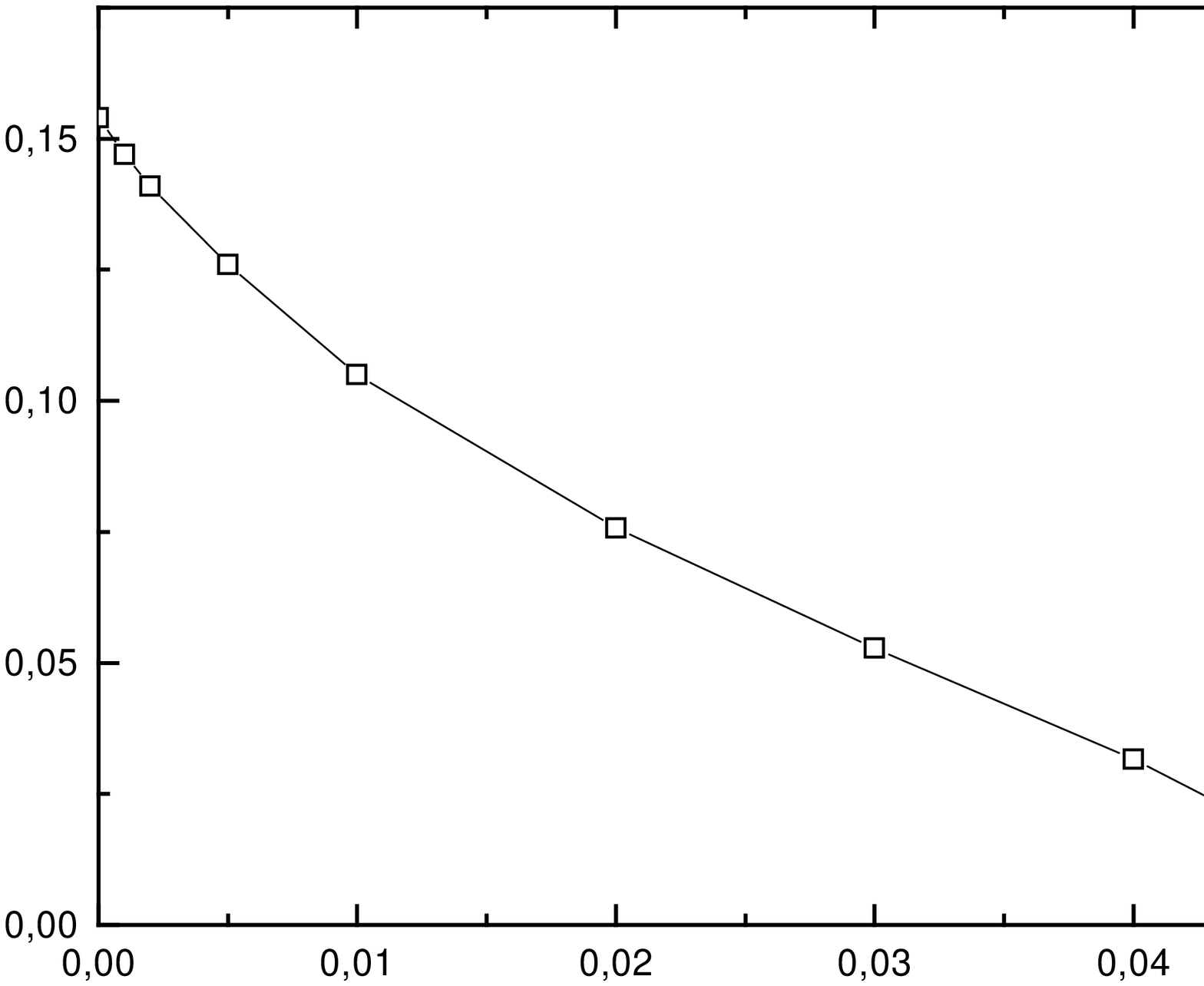, scale=0.6}
\end{figure}
\begin{figure}[htbp]
\epsfig{figure=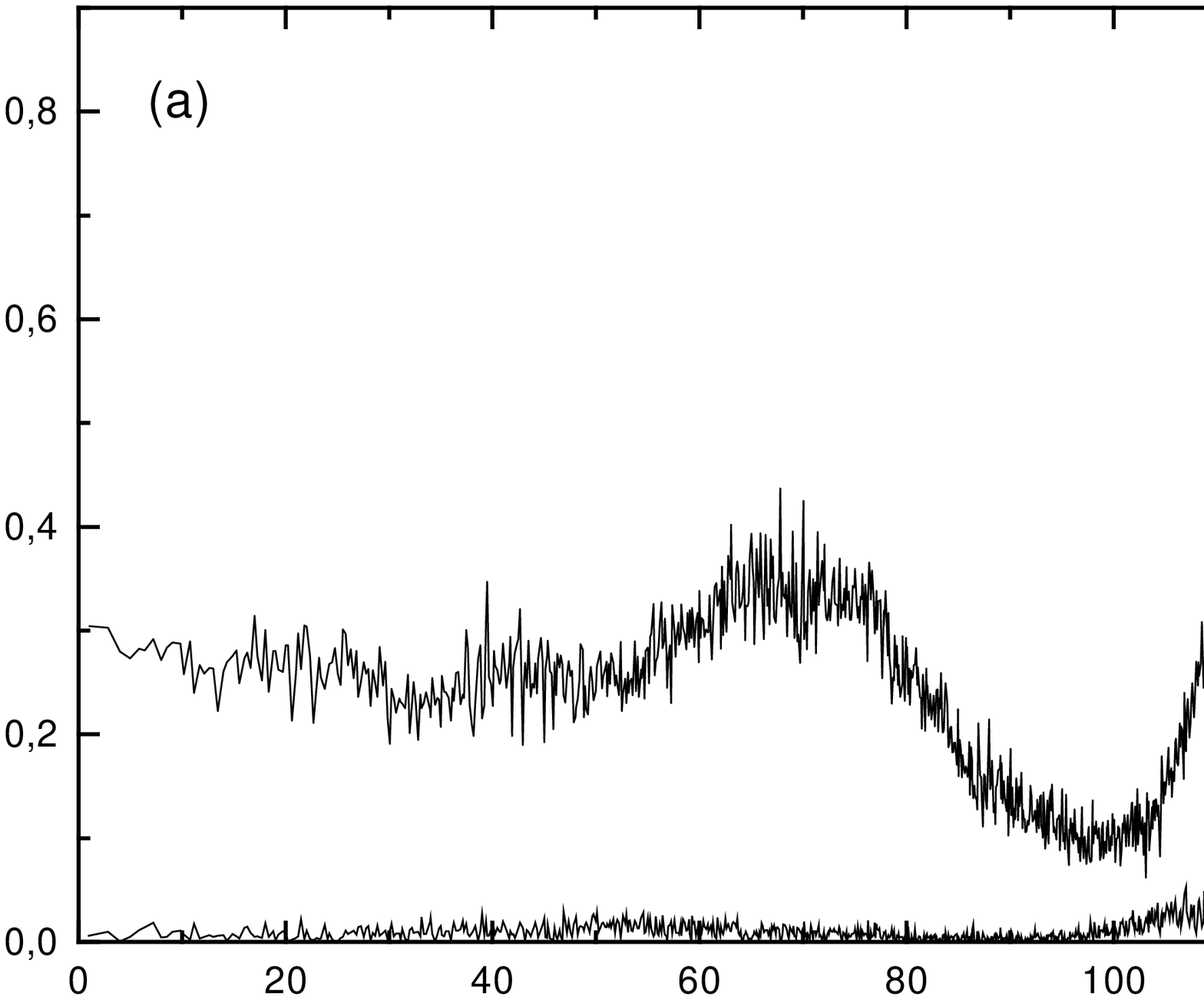, scale=0.6}
\end{figure}
\begin{figure}[htbp]
\epsfig{figure=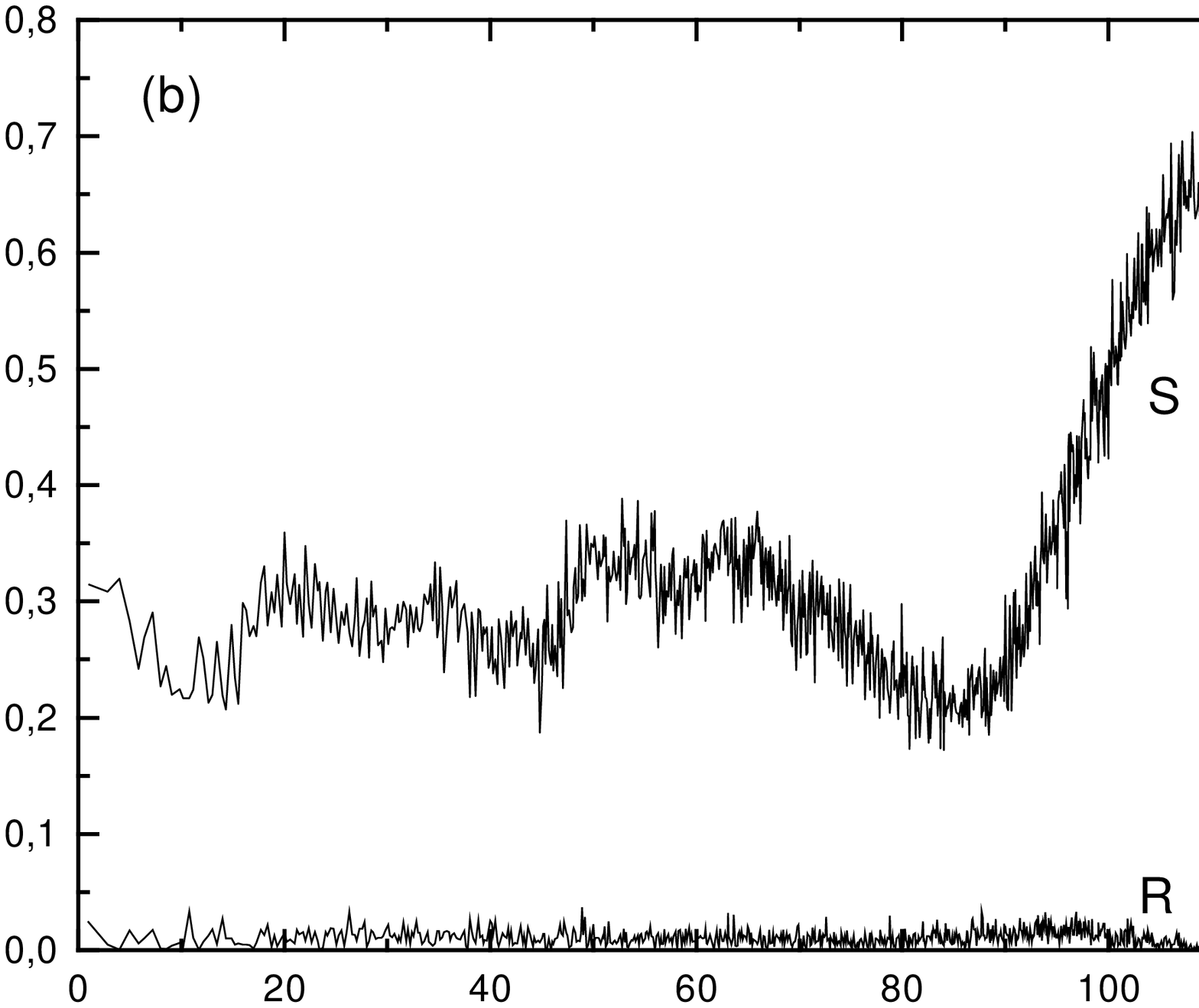, scale=0.6}
\end{figure}
\begin{figure}[htbp]
\epsfig{figure=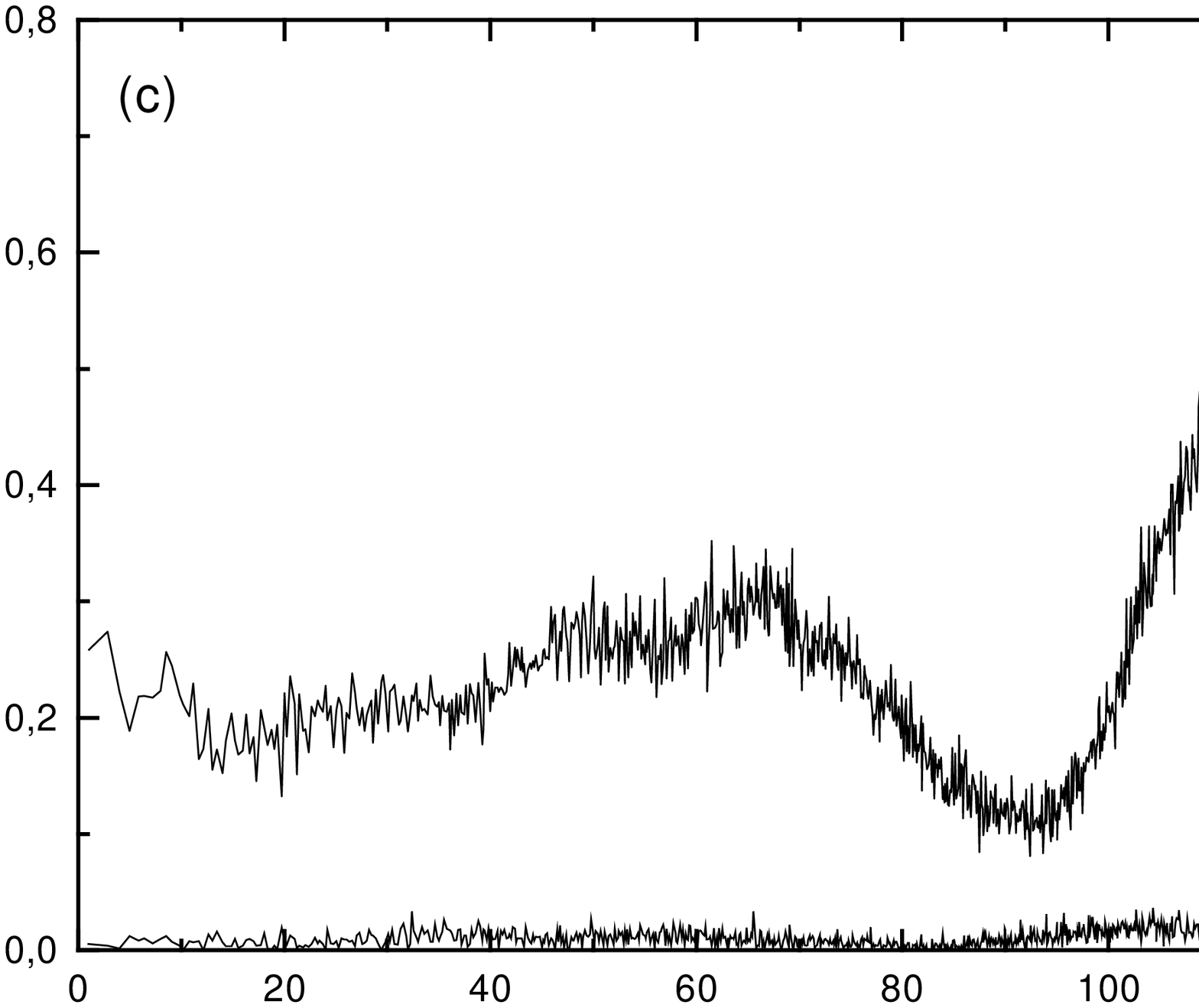, scale=0.6}
\end{figure}
\begin{figure}[htbp]
\epsfig{figure=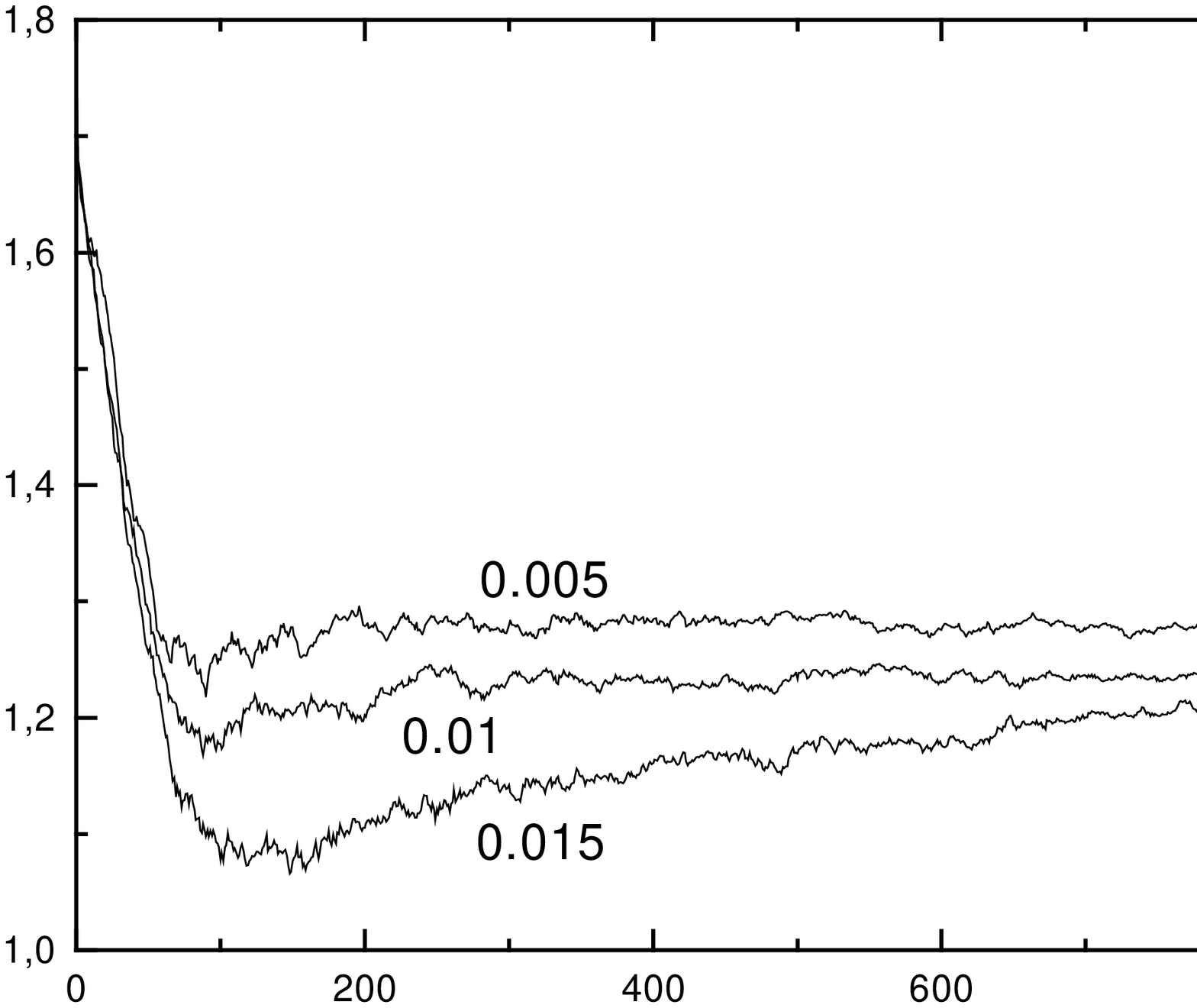, scale=0.6}
\end{figure}
\begin{figure}[htbp]
\epsfig{figure=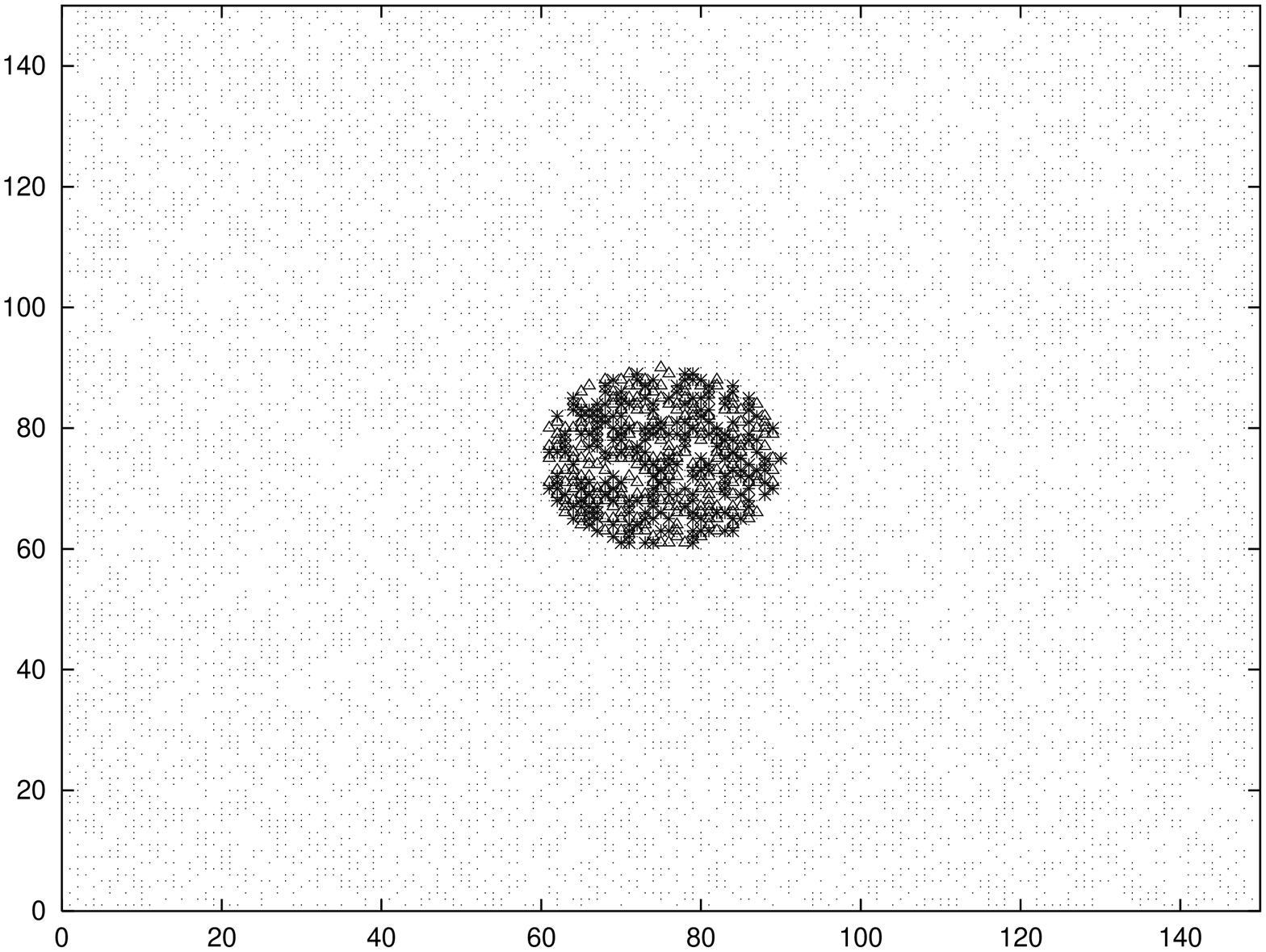, angle=-90, scale=0.5}
\end{figure}
\begin{figure}[htbp]
\epsfig{figure=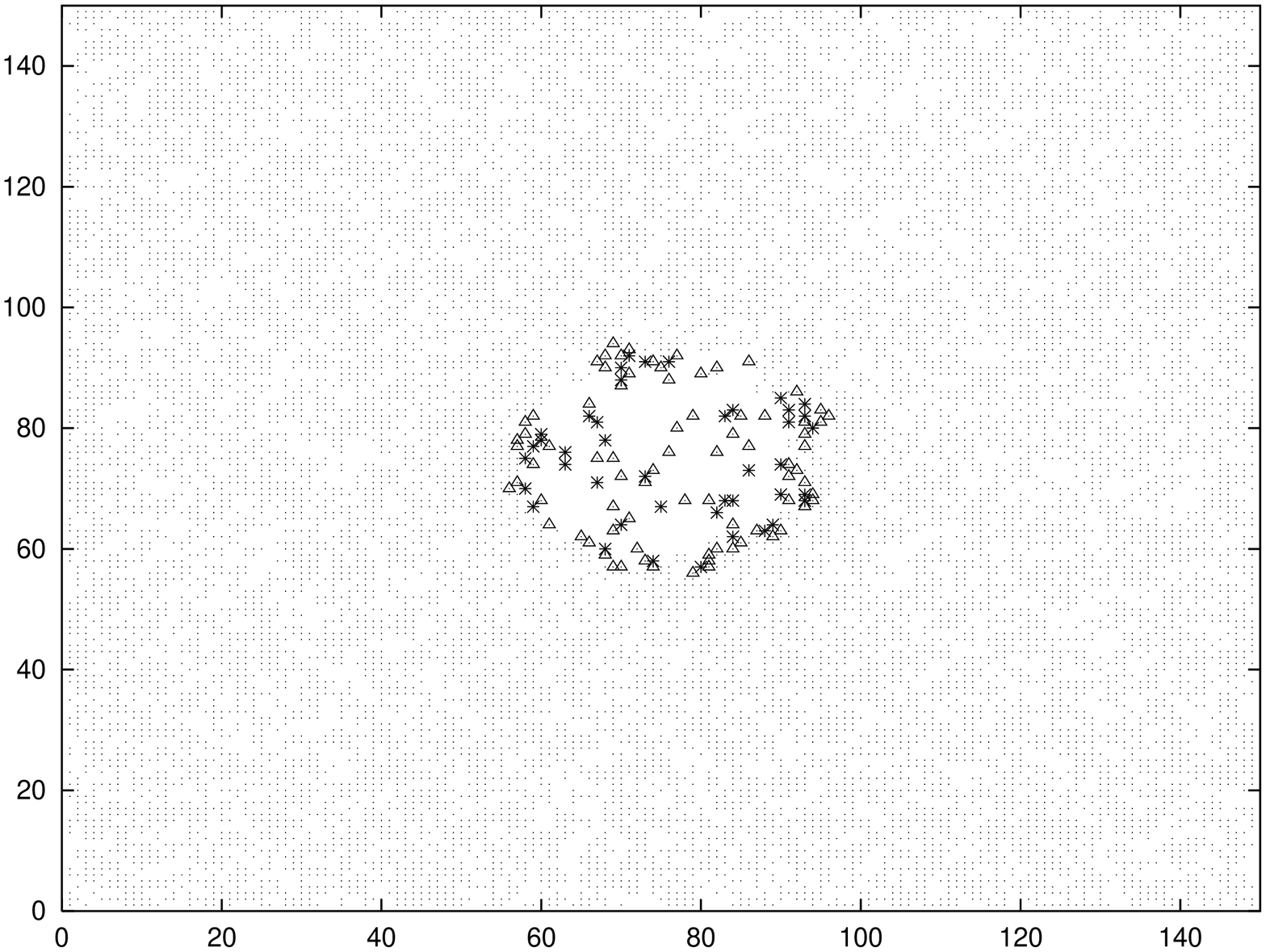, angle=-90, scale=0.5}
\end{figure}
\begin{figure}[htbp]
\epsfig{figure=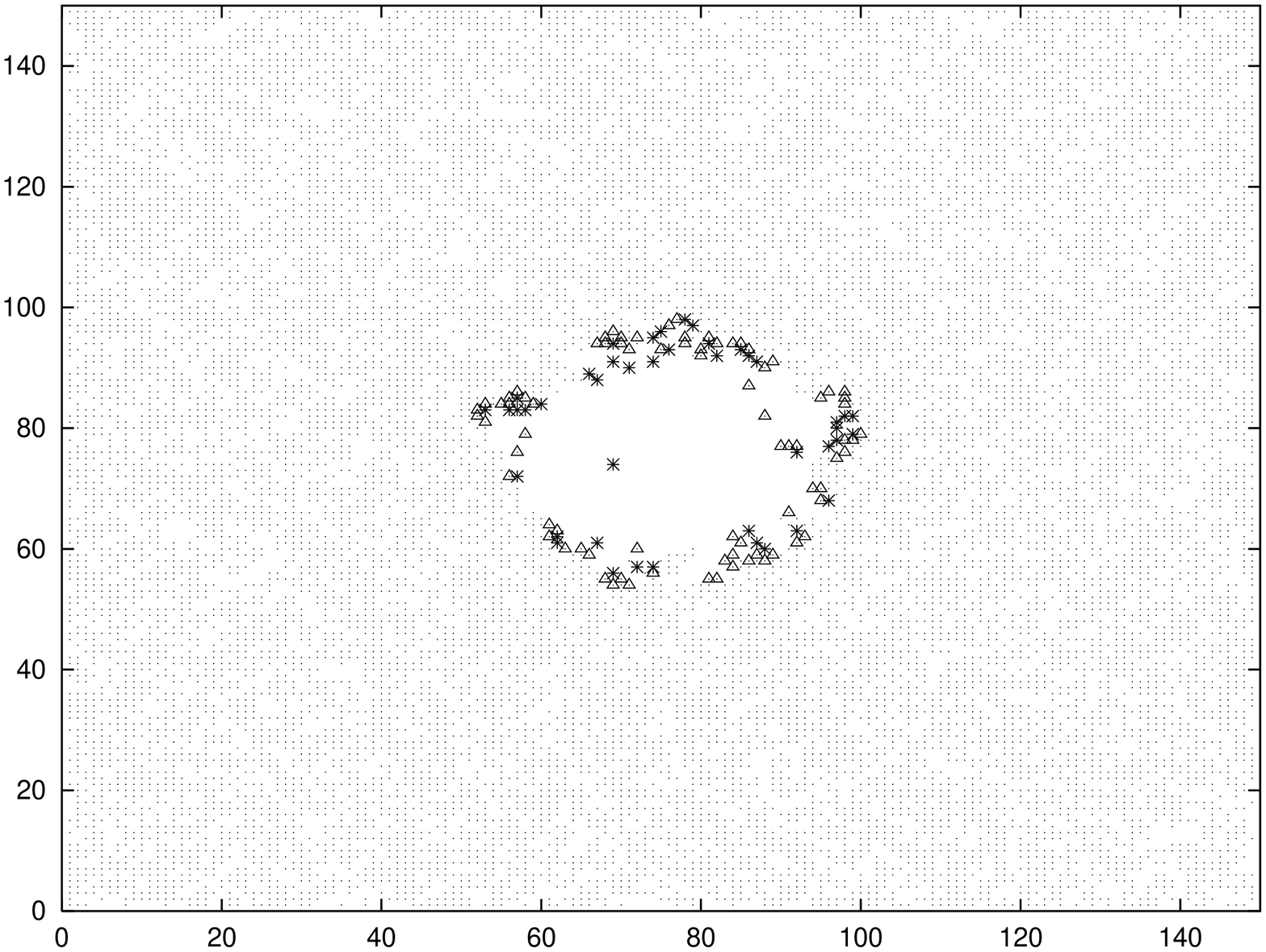, angle=-90, scale=0.5}
\end{figure}
\begin{figure}[htbp]
\epsfig{figure=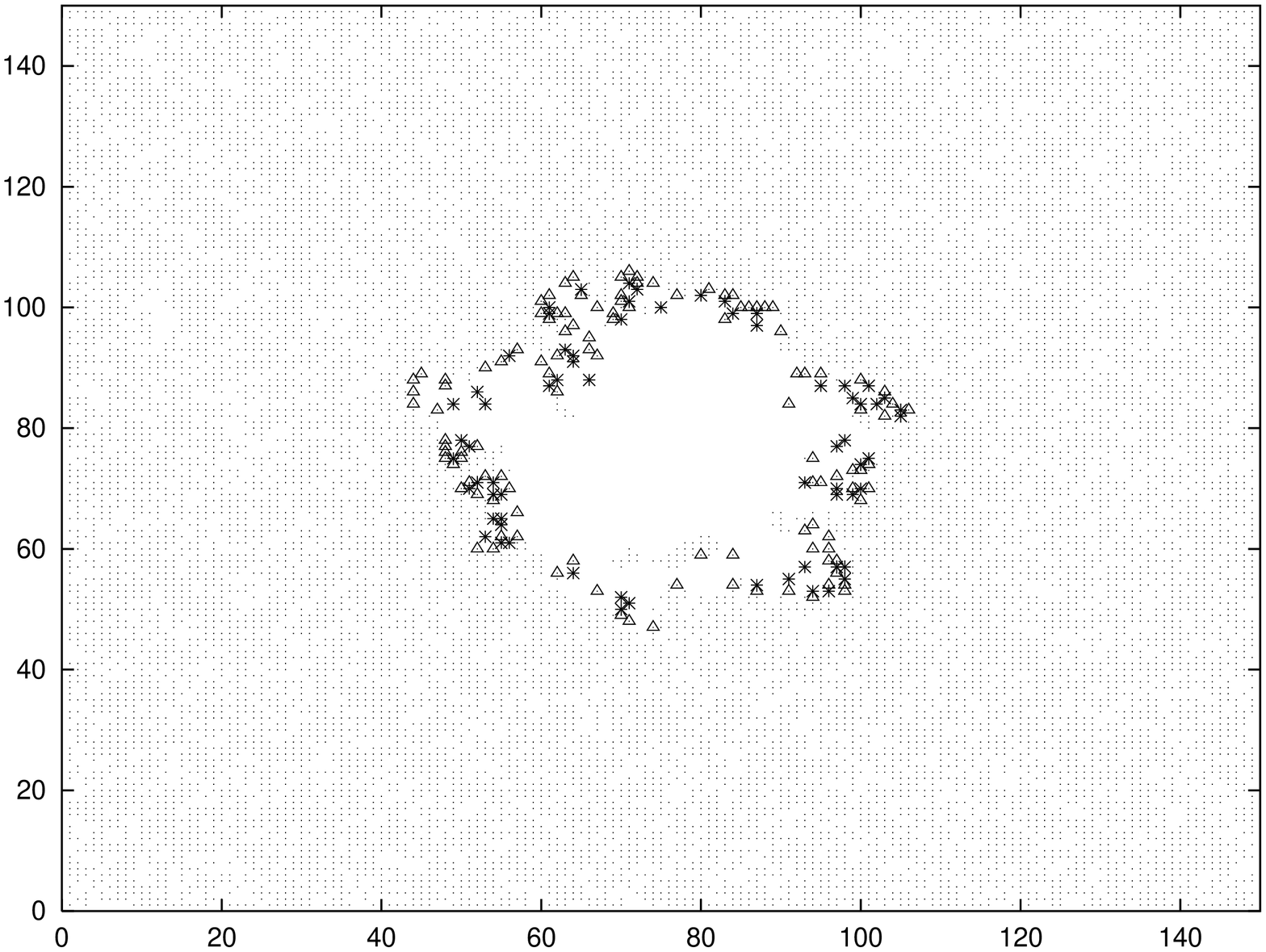, angle=-90, scale=0.5}
\end{figure}
\begin{figure}[htbp]
\epsfig{figure=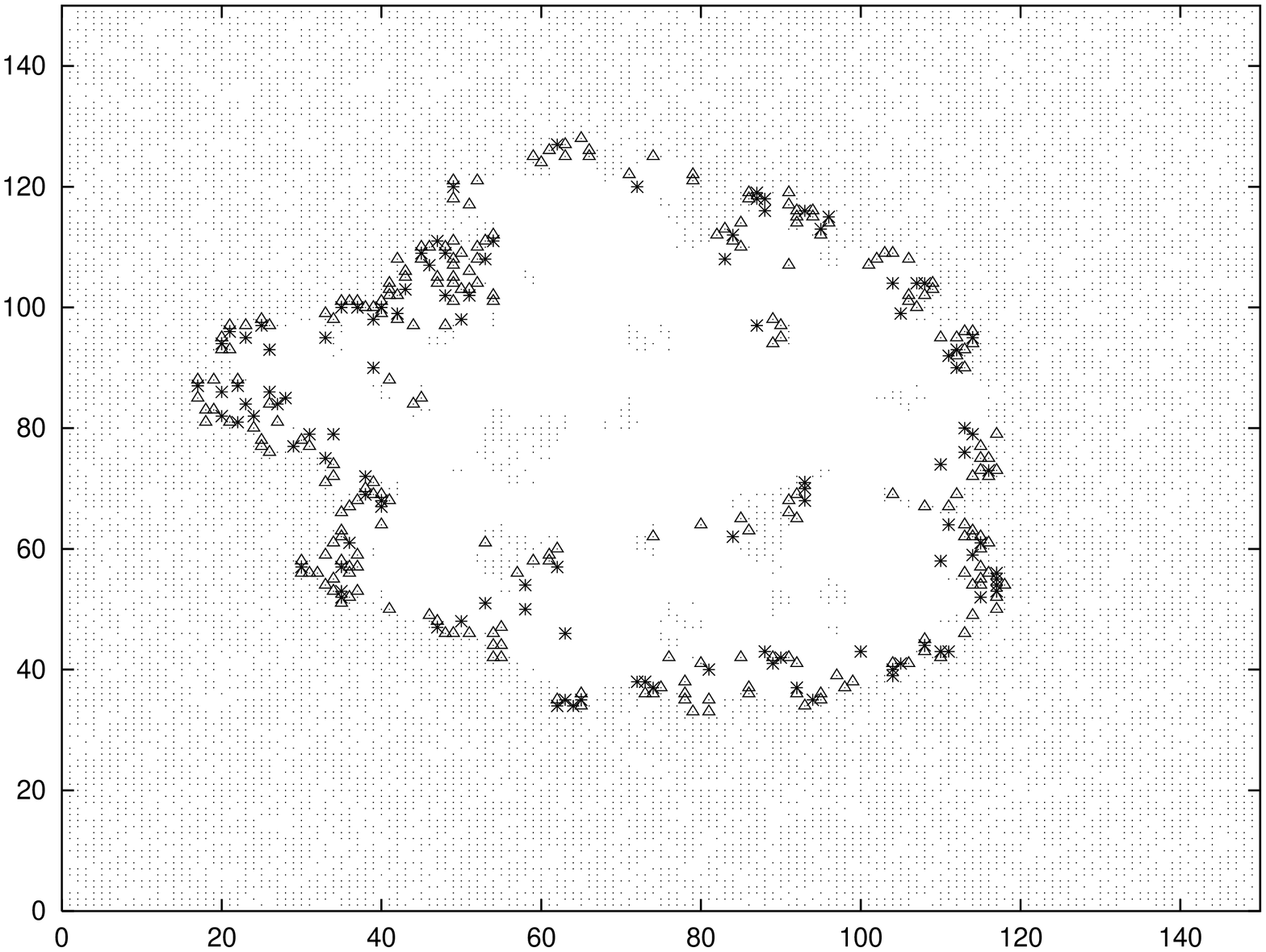, angle=-90, scale=0.5}
\end{figure}
\begin{figure}[htbp]
\epsfig{figure=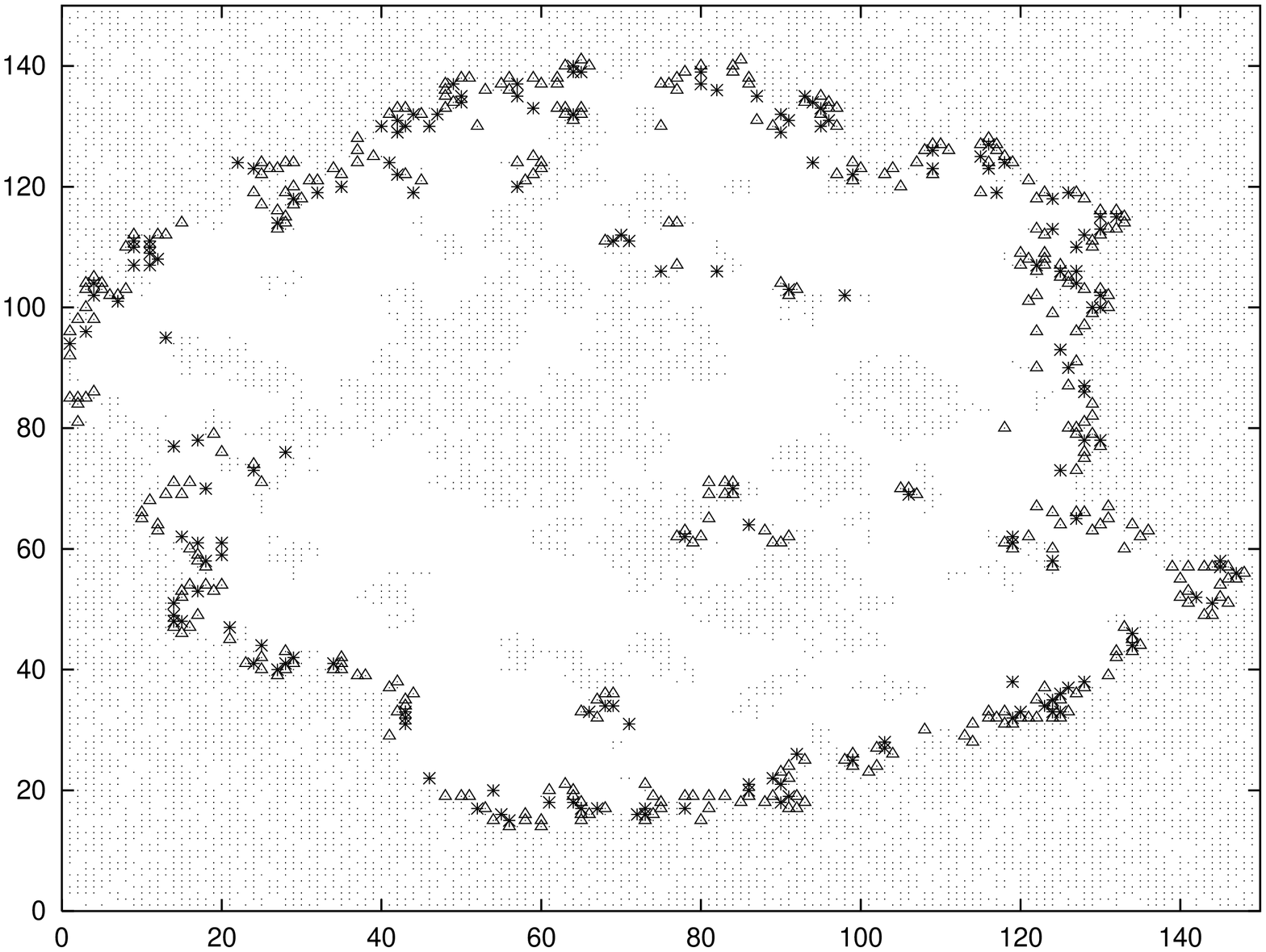, angle=-90, scale=0.5}
\end{figure}
\begin{figure}[htbp]
\epsfig{figure=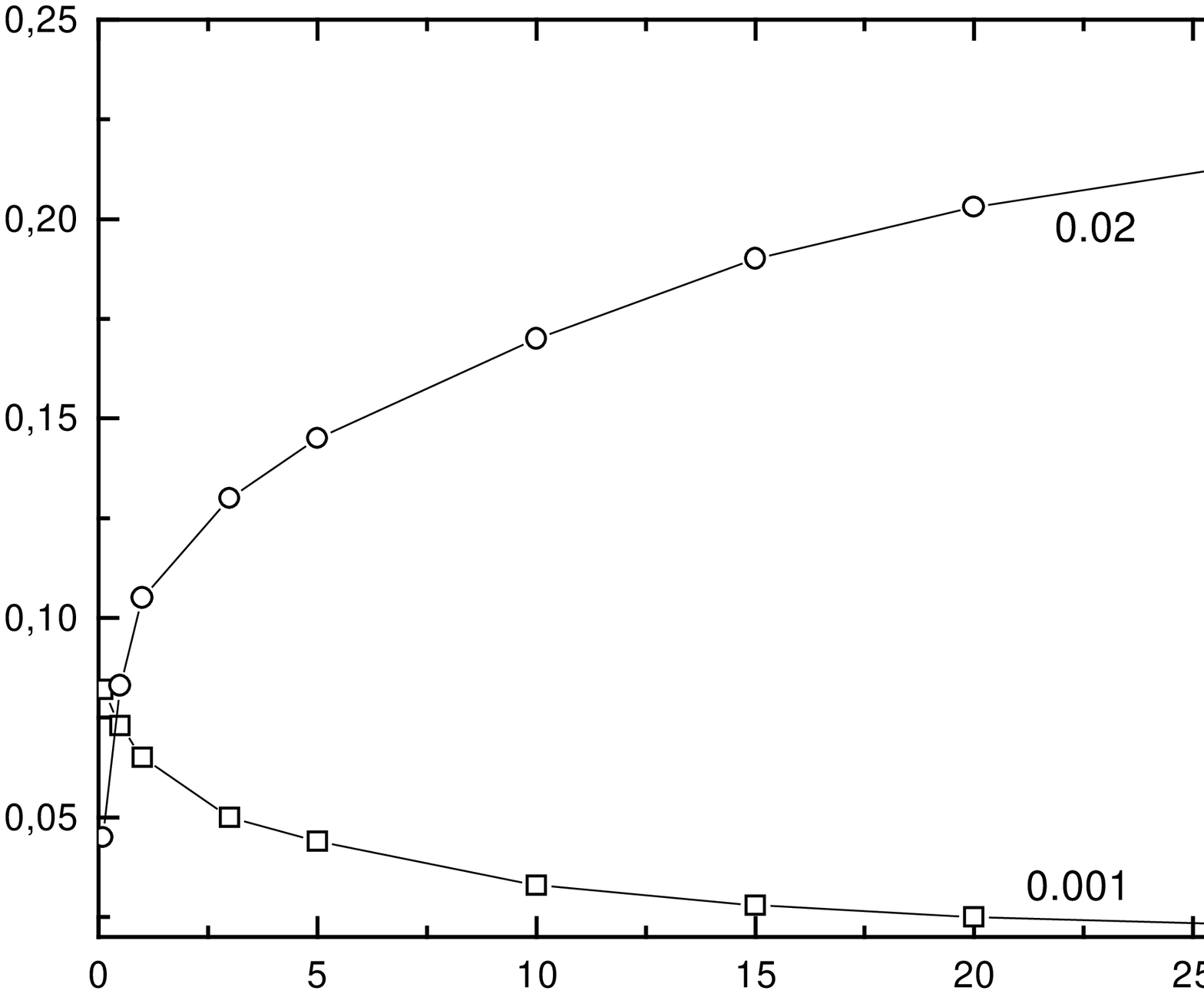, scale=0.6}
\end{figure}
\begin{figure}[htbp]
\epsfig{figure=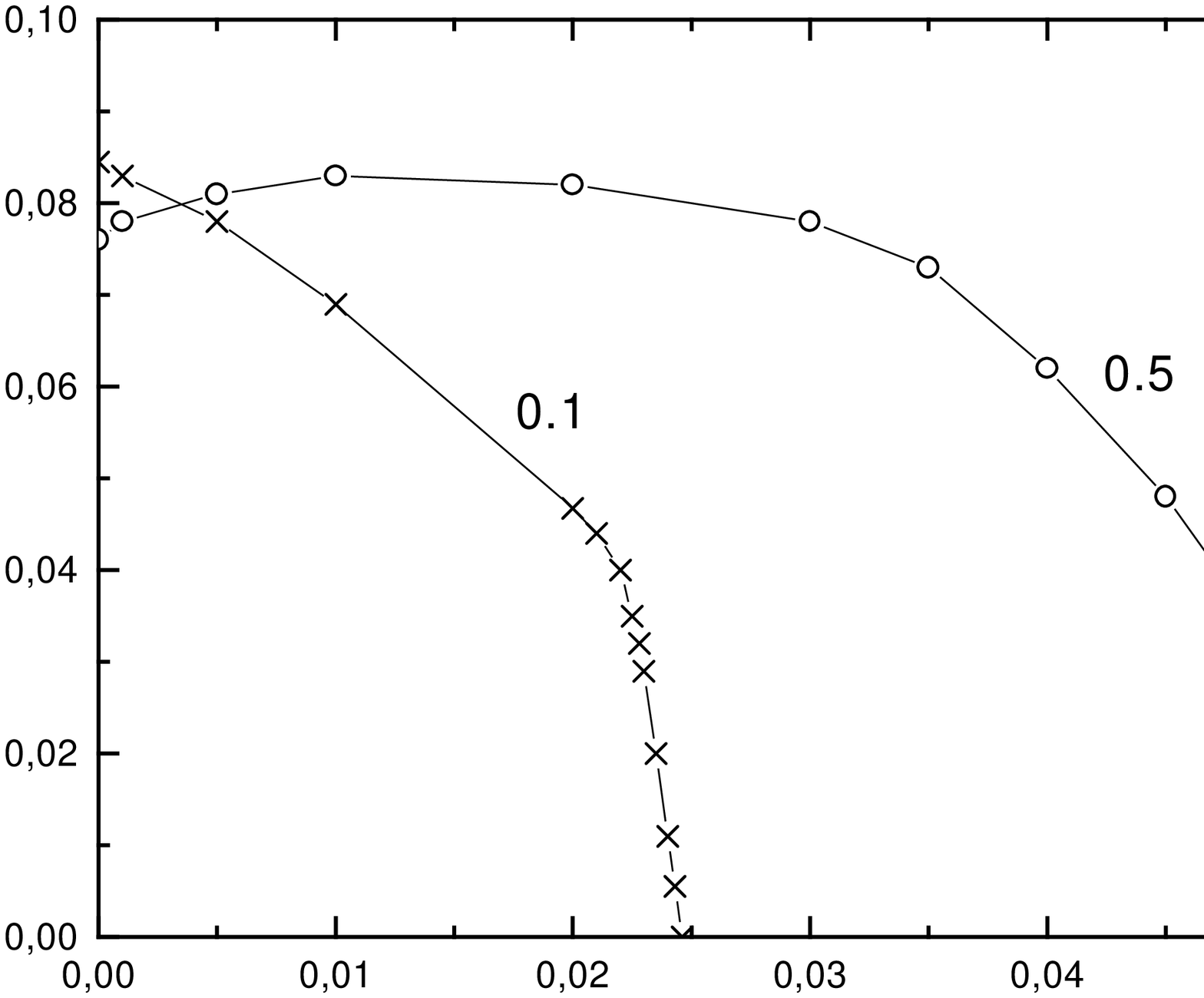, scale=0.6}
\end{figure}

\end{document}